\documentclass[nohyper,11pt,letterpaper]{JHEP3}

\pdfoutput=1
\usepackage{amssymb,amsmath}
\usepackage{epsfig}
\usepackage{color}
\usepackage{datetime}

\newcommand\fverb{\setbox\fverbbox=\hbox\bgroup\verb}
\newcommand\fverbdo{\egroup\medskip\noindent%
			\fbox{\unhbox\fverbbox}\ }
\newcommand\fverbit{\egroup\item[\fbox{\unhbox\fverbbox}]}
\newbox\fverbbox

\newcommand{\be}{\begin{equation}}
\newcommand{\ee}{\end{equation}}
\newcommand{\ba}{\begin{eqnarray}}
\newcommand{\ea}{\end{eqnarray}}

\newcommand{\pa}{\partial}

\newcommand{\lap}{\bigtriangleup}

\newcommand{\eq}[1]{(\ref{#1})}

\newcommand{\bs}[1]{{\boldsymbol{#1}}}

\newcommand{\hh}{\, ,\hspace{1cm}}
\newcommand{\hhh}{\, ,\hspace{0.4cm}}

\newcommand{\ins}[1]{{\mbox{\tiny #1}}}

\newcommand{\inds}[1]{{\scriptscriptstyle #1}}

\title{Biconformal symmetry and static Maxwell fields near
higher-dimensional black holes}

\author{Valeri P. Frolov\thanks{E-mail:
vfrolov@ualberta.ca}\, and
Andrei Zelnikov\thanks{E-mail: zelnikov@ualberta.ca}\\
Theoretical Physics Institute, Department of Physics,
University of Alberta, Edmonton, AB, Canada T6G 2E1}

\abstract{
We study an electric field created by a static electric charge near the 
higher dimensional Reissner-Nordstr\"{o}m  black hole. The relation between the static Green 
functions on the $D$-dimensional Reissner-Nordstr\"{o}m background and on the $(D+2)$-dimensional 
homogeneous Bertotti-Robinson spacetime is found. Using the biconformal symmetry we obtained a 
simple integral representation for the static Maxwell Green functions in arbitrary dimensions. 
We show that in a four-dimensional spacetime the static Green function obtained by the biconformal 
method correctly reproduces known results. We also found a closed form for the exact static Green 
functions and vector potentials in the five-dimensional  Reissner-Nordstr\"{o}m 
spacetime. 
}

\keywords{conformal symmetry, black holes, higher dimensions}
\preprint{Alberta Thy 5-15}

\begin{document}

\section{Introduction}

In this paper we continue studying fields created by static
charges placed in the vicinity of a higher-dimensional static black hole. For this purpose we use
the method of biconformal transformations, which was developed in our previous
paper \cite{Frolov:2014kia,Frolov:2014fya} in application to the case of scalar charges
in the Schwarzschild-Tangherlini and the Reissner-Nordstr\"{o}m geometries.

A vector potential $A_{\mu}$ in a  $D$-dimensional spacetime with metric $g_{\mu\nu}$
($\mu,\nu=0,\ldots, D-1$) obeys the Maxwell equations
\be\label{Maxwell}
F^{\mu\nu}{}\!_{;\nu}=4\pi J^{\mu} \hh F_{\mu\nu}=\partial_\mu A_\nu-\partial_\nu
A_\mu .
\ee
Here coefficient $4\pi$ comes from the definition of the Maxwell action in higher dimensions. In 
four dimensions our choice of units corresponds to the 
conventional Gaussian units. 
There is an ambiguity in generalizations of the Maxwell equations to higher dimensions which 
depends 
on system of units and the definition of an electric charge. In this paper charges are
normalized in such a way that the interaction force between two charges in $D$ dimensions reads
\be\label{f}
f={4\pi\over\Omega_\ins{(D-2)}}\,{e_1 e_2\over r^{D-3}}\hh
\Omega_n={2\pi^{(n+1)/2}\over \Gamma((n+1)/2)}.
\ee
Different choices of units in higher dimensions are also used in the literature. For example, in 
the paper \cite{Beach:2014aba} authors work in a different system of
units, such that the force
between two charges $\tilde{e}_1$ and $\tilde{e}_2$ in the $D-$dimensional
Minkowski spacetime is given by
\be
f={\tilde{e}_1 \tilde{e}_2\over r^{D-3}}\,.
\ee
Thus, our normalization of the charges and that of the paper
\cite{Beach:2014aba} are related as
\be
e^2={\Omega_{(D-2)}\over 4\pi}~ \tilde{e}^2 ,
\ee 
where the volume $\Omega_n$ of $n$-dimensional sphere $S^n$ is given in \eq{f}. In particular one 
has
\be
\Omega_2=4\pi\hhh
\Omega_3=2\pi^2\hhh
\Omega_4={8\pi^2\over 3}\hhh
\Omega_5=\pi^3
.
\ee

In the Lorentz gauge $\nabla^{\mu}A_{\mu}=0$  the Maxwell equations become
\be
\Box A_{\mu}-R_{\mu}^{\nu}A_{\nu}=-4\pi J_{\mu}.
\ee
Let us consider the potential created by a static electric source
$J^{\mu}=j(x)\,\delta^{\mu}_0$, where $j(x)$ is the charge
density in a static spacetime described by the metric
\be\begin{split}\label{met}
&ds^2=-\alpha^2 dt^2+g_{ab}\,dx^a dx^b, \\
&X^{\mu}=(t,x^a)\hhh \alpha=\alpha(x)\hhh g_{ab}=g_{ab}(x)\hh a=1,\dots, D-1\,.
\end{split}\ee
In the static case one can choose $A_a=0$ and $A_0=A_0(x)$, that is
\be
A_\mu=A_0\,\delta^0_\mu.
\ee
Then the Maxwell equations (\ref{Maxwell}) for the potential $A_\mu$ boil down to
\be\label{eqF}
\hat{O}\,A_0=4\pi j\hh
\hat{O}={1\over\alpha\sqrt{g}}\partial_a \left({1\over \alpha}\sqrt{g}
g^{ab}\partial_b\right).
\ee
Here $g=\det(g_{ab})$. The red-shift factor $\alpha$
 is connected with the norm of the static Killing vector $\bs{\xi}$ as follows:
$\alpha=\sqrt{-\bs{\xi}^2}=\sqrt{-g_{tt}}$.

We define the static Green function for the operator $\hat{O}$ as the solution of the following 
equation
\be\label{G00}
\hat{O}\,G_{00}(x,x')={1\over \alpha\sqrt{g}}\delta(x-x').
\ee
The equation (\ref{eqF}) is
invariant under the following {\em biconformal} transformations
\be\label{bi-conf}
A_0=\bar{A}_0 \hh g_{ab}=\Omega^2 \,\bar{g}_{ab}\hh
\alpha=\Omega^{n}\bar{\alpha}\hh
j=\Omega^{-2n-2}\,\bar{j},
\ee
where $n\equiv D-3$ and $\Omega$ is an arbitrary function of spatial coordinates
$x^a$. 

These biconformal transformations can be used to relate solutions of the Maxwell equations on 
a physical metric to solutions on a some other `reference' geometry. If the reference spacetime is 
more symmetrical than the original one, then there is a good chance to simplify the problem of 
finding the Green function exactly. This approach, for example, enabled us \cite{Frolov:2012jj} to 
compute the Green functions of static scalar and Maxwell fields on the background of the  
Majumdar-Papapetrou spacetime, which describe a set of extremally charged black holes. It was 
possible because the higher dimensional Majumdar-Papapetrou metric is biconformally related to the 
flat Minkowski metric. 

In this paper we use biconformal transformations to compute the static Green functions of the 
Maxwell field on the background of a generic higher dimensional Reissner-Nordstr\"{o}m  black hole.


\section{Point charge near higher-dimensional Reissner-Nordstr\"{o}m black hole}

Let us consider a 
static spherically symmetric $D$-dimensional metric of the form
\be\label{ds}
ds^2=-f(r)\,dt^2+f^{-1}(r)\,dr^2+r^2\,d\omega^2_{n+1}\, ,
\ee
where $n=D-3$ and  $d\omega_{n+1}^2$ is the line
element on a $(n+1)$-dimensional unit sphere
\be\label{theta}
d\omega_{n+1}^2=d\theta_{n}^2+\sin^2\theta_{n}\,d\omega_{n}^2\hh
d\omega_{0}^2 = d\phi^2\hh
\theta_n=\theta .
\ee
We denote $\theta_0\equiv\phi\in[0,2\pi]$. The other angular coordinates
$\theta_{i>0}\in[0,\pi]$.

\be\label{RN}
f=1-{2M\over r^n}+{Q^2\over r^{2n}} .
\ee
For real positive $M$ and real $Q$, which satisfy the condition $|Q|\le M$, the metric \eq{ds}
-\eq{RN} describes the geometry of a higher dimensional generalization of a spherically 
symmetric
electrically charged black hole\footnote{Here we consider an electric field of a test charge on 
the background of an electrically charged black hole. A total electric field is the sum of the 
electric field of test charge and of the black hole. This distinction becomes important, when one 
considers quantities, which are nonlinear in the field, for example, an energy of the Maxwell field 
or a self-force of a charge.}. 
The parameters $M$ and $Q$ are proportional to the 
Arnowitt-Deser-Misner mass and charge of the black hole, respectively. The coefficients of 
proportionality (see, e.g., 
\cite{Chamblin:1999hg}) depend on the dimensionality of the spacetime and on the choice of units.

It is convenient to introduce a new radial variable $\rho$ related to the
radial coordinate $r$ as follows
\be\label{rho_r}
\rho={r^n -M\over \mu}\hh \mu=\sqrt{M^2-Q^2}\hh r^n=M+\mu\rho\, .
\ee
Then the Reissner-Nordstr\"{o}m metric \eq{ds}-\eq{RN} takes the form
\be\begin{split}\label{R-N}
ds^2&=-{\mu^2(\rho^2-1)\over(M+\mu\rho)^2}\,dt^2
+(M+\mu\rho)^{2/n}\left[{1\over
n^2(\rho^2-1)}\,d\rho^2+d\omega^2_{n+1} \right] .
\end{split}\ee
The horizon corresponds to
$
\rho=1,
$
and its (gravitational) radius $r_\ins{g}$ is given by the expression
$
r_\ins{g}^n=M+\mu .
$
The surface gravity at the horizon is
\be\label{kappa}
\kappa={n\mu \over r_\ins{g}^{n+1}} .
\ee

Taking into account that
\be\begin{split}
&\alpha={\mu\sqrt{\rho^2-1}\over M+\mu\rho}\hhh g^{\rho\rho}=n^2(\rho^2-1)(M+\mu\rho)^{-2/n}
\hhh
g^{\theta\theta}=(M+\mu\rho)^{-2/n},\\
&\sqrt{-g^\ins{D}}=\alpha\sqrt{g}={1\over n}{\mu\,(M+\mu\rho)^{2/n}}\sqrt{g_\omega}
\hh
\sqrt{g_\omega}=\prod_{k=1}^{n}(\sin\theta_k)^k,
\end{split}\ee
the equation for the Green function \eq{G00} takes the form
\be
\left[n^2\,\partial_{\rho}\,(M+\mu\rho)^2\,\partial_{\rho}+{(M+\mu\rho)^2\over\rho^2-1}\lap_{\omega}
^ { n+1} \right]G_{00}(x,x')=n\mu\,\delta(\rho-\rho')\delta(\omega,\omega') .
\ee
If we use the ansatz
\be
G_{00}(x,x')=-{\mu^2\,(\rho^2-1)(\rho'{}^2-1)\over (M+\mu\rho)(M+\mu\rho')}\,H(x,x') ,
\ee
then the equation for $H(x,x')$ becomes
\be
\left\{n^2[(\rho^2-1)\,\partial^2_{\rho}+4\rho\,\partial_{\rho}+2]+\lap^{n+1}_{\omega}\right\}\,H(x,
x')=-{n\over\mu (\rho'{}^2-1)}\,\delta(\rho-\rho')\delta(\omega,\omega') .
\ee

In order to solve this equation we shall use the following trick. We first consider another
equation for the static Green function of a massive scalar
operator $\Box-m^2$ with the mass
\be\label{m2}
m^2=-2n^2/a^2=-2/b^2
\ee
on a $(D+2)$-dimensional homogeneous spacetime, which is a direct product
of the $(n+1)$-dimensional sphere of a radius $a$ and a four-dimensional AdS spacetime 
or, in the Euclidean version, the hyperboloid
$H^4$ of a radius $b=a/n$.
\be\label{H4S}
d\tilde{s}^2=d\ell^2_\inds{H^4}+a^2\,d\omega^2_{n+1}\hh b={a\over n}\hh 
X_\ins{E}=(\rho,\sigma,\bar{\theta},\bar{\phi},\theta, \theta_{n-1},\dots,\phi),
\ee
\be\label{dell2}
d\ell^2_\inds{H^4}=b^2\left[{1\over \rho^2-1}\,d\rho^2+(\rho^2-1)\,d\bar{\omega}^2_{3}\right] ,
\ee
\be\label{bar_omega}
d\bar{\omega}_{n+1}^2=d\bar{\theta}_{n}^2+\sin^2\bar{\theta}_{n}\,d\bar{\omega}_{n}^2\hh
d\bar{\omega}_{0}^2 = d\bar{\phi}^2\hh
\bar{\theta}_n=\sigma .
\ee

The Green function of an Euclidean massive operator 
\be\label{hatO}
\hat{O}=\Box_\ins{E} -m^2
\ee
defined on a $(D+2)$-dimensional Bertotti-Robinson spacetime, can be obtained by the heat 
kernel method. In order to calculate the function 
$H(x,x')$ one has to find at first the $(D+2)$-dimensional Green function of the operator 
\eq{hatO}, 
which satisfies the equation
\be
\hat{O}\,{\mathbb G}_\inds{\hat{O}} =-\delta^\inds{D+2}(X_\ins{E},X'_\ins{E}).
\ee
The hyperboloid $H^4$ is spherically symmetric. We shall show that integration of ${\mathbb 
G}_\inds{\hat{O}}$  
over all angle coordinates $\bar{\theta}_{k}$ on $H^4$ gives 
\be
H(x,x')={a^{n+3}\over  n^3\mu}\int
d^3\bar{\omega}\sqrt{g_\inds{\bar{\omega}}}\,{\mathbb G}_\inds{\hat{O}}(X_\ins{E},X'_\ins{E}) .
\ee
This approach is similar to the case of a scalar field
\cite{Frolov:2014kia,Frolov:2014fya}.
The difference is that the required static Green function in the spacetime of a black hole is 
generated by the Green function of the massive scalar operator, which is defined on 
the $H^4\times S^{n+1}$ geometry.


\section{Green functions and heat kernels}

\subsection{General formulas}

The Euclidean Green
function for the self-adjoint operator $\hat{O}_\ins{E}$ in any dimensions can be written in terms
of the heat kernel of this operator
\be
{\mathbb 
G}_\inds{\hat{O}}(X_\ins{E},X'_\ins{E})=\int_0^{\infty}ds\,K_\inds{\hat{O}}(s|X_\ins{E},X'_\ins{E}
).
\ee
Here the heat kernel $K_\inds{\hat{O}}(s|X_\ins{E},X'_\ins{E})$ is the solution of the problem
\be
(\pa_s-\hat{O})\,K_\inds{\hat{O}}(s|X_\ins{E},X'_\ins{E})=0
\hh
K_\inds{\hat{O}}(0|X_\ins{E},X'_\ins{E})=\delta(X_\ins{E},X'_\ins{E})
\,,
\ee
which satisfies the same boundary conditions with respect to its arguments
$X_\ins{E}$
and $X'_\ins{E}$ as the Green function in question.

Because the geometry of the $(D+2)$-dimensional
Bertotti-Robinson spacetime has the form of a direct sum of two
homogeneous spaces, the heat  $K$ has a form of a
product of the heat kernels $K_\inds{H^4}$ and $K_\inds{S^{n+1}}$ for the reduced box 
operators defined on the
hyperboloid $H^4$ and on the sphere $S^{n+1}$, respectively
\be\begin{split}\label{KBR}
K_\inds{\hat{O}}(s|\rho,\bar{\theta}_k,\theta_n;\rho',\bar{\theta}'_k,\theta'_n)&=e^{-m^2s}\,K_\inds
{ H^4}(s|\rho,\bar{\theta}_k;\rho',\bar{\theta}'_k)\,K_\inds{ S^ { n+1 } }
(s|\theta_n;\theta'_n) .
\end{split}\ee
Both spaces $H^4$ and $S^{n+1}$ are
homogeneous and isotropic and the corresponding heat kernels are known
explicitly \cite{Camporesi:1990wm}.

In the case of the operator $\Box_\ins{E}-m^2$ defined on the $n+5$-dimensional Bertotti-Robinson
metric \eq{H4S} the Euclidean Green function satisfies the equation
\be
\left(\Box_\ins{E} -m^2\right){\mathbb
G}_\inds{\hat{O}}(X_\ins{E},X'_\ins{E})=-\delta(X_\ins{E},X'_\ins{E}) .
\ee
Because of the symmetry of the metric, this Green function is, in fact, the function of only two
geodesic distances: $\chi$ between points on the hyperboloid and $\gamma$ on the sphere 
\be
{\mathbb G}_\inds{\hat{O}}(X_\ins{E},X'_\ins{E})={\mathbb G}_\inds{\hat{O}}(\chi,\gamma) .
\ee
Explicitly, the equation for the $(D+2)$-dimensional Green function ${\mathbb
G}_\inds{\hat{O}}$ reads
\be\begin{split}
\left\{n^2\left[(\rho^2-1)\,\partial^2_{\rho}+4\rho\,\partial_{\rho}+{1\over\rho^2-1}\lap^{
3}_{ \bar{\omega} }\right]-a^2 m^2+\lap^{n+1}_{ \omega } \right\} \,{\mathbb
G}_\inds{\hat{O}}(\chi,\gamma)&\\
=-{n^4\over a^{n+3}
(\rho'{}^2-1)}\,\delta(\rho-\rho')\delta(\bar{\omega},\bar{\omega}')\delta(\omega, \omega') .&
\end{split}\ee


\subsection{Heat kernel on $H^4$}

The heat kernel of the four-dimensional Laplace operator defined on the hyperboloid \eq{dell2} of 
the radius $b$
reads
\be\label{KH4}
K_\inds{H^4}(s|\chi)=-e^{-2s/b^2}\left({1\over 2\pi
b^2}{\partial\over\partial\cosh\chi}\right)K_\inds{H^2}(s|\chi) .
\ee
Here $K_\inds{H^2}(s|\chi)$ is the heat kernel of the Laplace operator defined on the 
two-dimensional hyperboloid $H^2$ of the
same radius
\be\label{H2}
d\ell^2_\inds{H^2}=b^2\left[(\rho^2-1)^{-1}\,d\rho^2+(\rho^2-1)\,d\sigma^2\right] .
\ee
It reads \cite{Camporesi:1990wm}
\be
K_\inds{H^2}(s|\chi)={\sqrt{2} b \over (4\pi
s)^{3/2}}\,e^{-s/(4b^2)}\int_{\chi}^{\infty}dy\,{y\,e^{-b^2
y^2/(4s)}\over(\cosh y-\cosh(\chi))^{1/2}} .
\ee
Here $\chi$ is the geodesic distance between
two points on the $H^2$ of the unit radius $b=1$. It is given by the
relation
\be\begin{split}\label{chi}
\cosh(\chi)&=\rho\rho'-\sqrt{\rho^2-1}\sqrt{\rho'{}^2-1}\cos(\sigma-\sigma') .
\end{split}\ee


\subsection{Heat kernel on $S^{n+1}$}

The heat kernel on a two-dimensional sphere $S^2$ of the radius $a$ reads \cite{Camporesi:1990wm}
\be
K_\inds{S^2}(s|\gamma)={\sqrt{2} a \over (4\pi
s)^{3/2}}\,e^{s/(4 a^2)}\sum_{k=-\infty}^{\infty}(-1)^k
\int_{\gamma}^{\pi} d\phi {(\phi+2\pi
k)\,e^{-a^2(\phi+2\pi k)^2/(4s)}\over(\cos\gamma-\cos\phi)^{1/2}} .
\ee
Another equivalent representation of this kernel is
\be
K_\inds{S^2}(s|\gamma)={1\over 4\pi a^2}
\sum_{l=0}^{\infty}(2l+1)P_l(\cos\gamma)\,e^{-{sl(l+1)\over a^2}} .
\ee
Here $\gamma$ is the
geodesic distance between two points on the unit $S^2$ ($a=1$)
\be
\cos\gamma=\cos(\theta_1)\cos(\theta'_1)
+\sin(\theta_1)\sin(\theta'_1)\cos(\phi-\phi') .
\ee

The heat kernel on the three-dimensional sphere $S^3$ of the radius $a$ reads 
\cite{Camporesi:1990wm}
\be
K_\inds{S^3}(s|\gamma)={1\over (4\pi
s)^{3/2}}\,e^{s/a^2}\sum_{k=-\infty}^{\infty}{(\gamma+2\pi k)
\,e^{-a^2(\gamma+2\pi k)^2/(4s)}\over\sin\gamma} ,
\ee
where $\gamma$ is the geodesic distance between two points on the unit $S^3$
($a=1$)
\be
\cos\gamma=\cos(\theta_2)\cos(\theta'_2)+\sin(\theta_2)\sin(\theta'_2)[
\cos(\theta_1)\cos(\theta'_1)+\sin(\theta_1)\sin(\theta'_1)\cos(\phi-\phi')] .
\ee

The heat kernels on all higher-dimensional spheres $S^{n+1}$ can be derived from $K_\inds{S^2}$ and
$K_\inds{S^3}$ using the relations (see \cite{Camporesi:1990wm}
Eq.(8.12)-Eq(8.13))
\be\label{KS}
K_\inds{S^{n+1}}(s|\gamma)=e^{(n^2-1)s\over 4 a^2}\left({1\over
2\pi a^2}{\partial\over \partial \cos\gamma}\right)^{(n-1)\over
2}K_\inds{S^2}(s|\gamma)\hh n~~\mbox{odd}  ,
\ee
\be
K_\inds{S^{n+1}}(s|\gamma)=e^{(n^2-4) s\over 4 a^2}\left({1\over
2\pi a^2}{\partial\over \partial \cos\gamma}\right)^{(n-2)\over
2}K_\inds{S^3}(s|\gamma)\hh n~~\mbox{even}  .
\ee


\subsection{Heat kernel and Green functions on $H^4\times S^{n+1}$}


For computational reasons, we also use another Green function, which is the Green function 
of the Laplace operator defined on the $D$-dimensional Euclidean   
Bertotti-Robinson space $H^2\times S^{n+1}$.
\be\label{H2S}
d\bar{s}^2=d\ell^2_\inds{H^2}+a^2\,d\omega^2_{n+1}\hh b={a\over n},
\ee
\be\label{dell}
d\ell^2_\inds{H^2}=b^2\left[{1\over \rho^2-1}\,d\rho^2+(\rho^2-1)\,d\sigma^2\right].
\ee
In the latter case the corresponding Euclidean Green function, which we denote $\bar{\mathbb
G}$, satisfies the equation
\be
\bar{\Box}_\ins{E}(X,X')\,\bar{\mathbb G}=-\delta(X,X') \hh 
X^\alpha=(\rho,\sigma,\theta_n,\dots,\phi).
\ee
Because of the symmetries of the Bertotti-Robinson spacetime, the Green function 
$\bar{\mathbb G}$ and the heat kernel $\bar{K}$ of the operator $\bar{\Box}_\ins{E}$ are 
functions of only the geodesic distances $\gamma$ and $\chi$ between the points on the sphere and on 
the hyperboloid, respectively. One can write explicitly
\be\begin{split}
\left\{n^2\left[(\rho^2-1)\,\partial^2_{\rho}+2\rho\,\partial_{\rho}+{1\over\rho^2-1}\,
\partial^2_ { \sigma}\right]+\lap^{n+1}_{ \omega } \right\}
\,\bar{\mathbb G}(\chi,\gamma)&\\
=-{n^2\over a^{n+1}}\,\delta(\rho-\rho')\delta(\sigma-\sigma')\delta(\omega, \omega') .&
\end{split}\ee

Using \eq{m2},\eq{KH4}, \eq{KS}, and \eq{KBR} one can express the heat 
kernel for the $(D+2)$-dimensional massive scalar operator $\hat{O}$ in terms of that of the 
$D$-dimensional massless operator $\bar{\Box}_\ins{E}$
\be
K_\inds{\hat{O}}(s|\chi,\gamma)=-\left({n^2\over 2\pi
a^2}{\partial\over\partial\cosh\chi}\right) \bar{K}(s|\chi,\gamma) ,
\ee
where we put $b=a/n$.
The heat kernel $\bar{K}(s|\chi,\gamma)$ is that of the massless scalar
Euclidean D'Alembert operator. It has been calculated in \cite{Frolov:2014kia,Frolov:2014fya}.

The Green function of the scalar operator is the integral
over the proper time $s$ of the corresponding heat kernel
\be
{\mathbb G}_\inds{\hat{O}}(\chi,\gamma)=\int_0^{\infty}ds\,K_\inds{\hat{O}}(s|\chi,\gamma) .
\ee
Therefore, one can write
\be
{\mathbb G}_\inds{\hat{O}}(\chi,\gamma)=-\left({n^2\over 2\pi
a^2}{\partial\over\partial\cosh\chi}\right) \bar{\mathbb G}(\chi,\gamma) .
\ee
Note that in this relation both ${\mathbb G}_\inds{\hat{O}}$ and $\bar{\mathbb G}$ are considered 
as functions of $\chi$ and $\gamma$. On the other hand one has to keep in mind that the geodesic
distances $\chi$ on $H^4$ and $H^2$ are quite different functions of coordinates on these 
hyperboloids. One should also remember the corresponding static Green functions are 
defined as integrals of ${\mathbb G}_\inds{\hat{O}}$ and $\bar{\mathbb G}$ over the Euclidean time 
$\sigma$ with different measures. The scalar Green function $\bar{\mathbb G}(\chi,\gamma)$ has been 
calculated in our
papers \cite{Frolov:2014kia,Frolov:2014fya}.

Taking into account that
\be
\int d^3\bar{\omega}\sqrt{g_\inds{\bar{\omega}}}\,\delta(\bar{\omega},\bar{\omega}') =1 \hh
\int d^3\bar{\omega}\sqrt{g_\inds{\bar{\omega}}}\,\lap^{3
}_{ \bar{\omega} } (\dots) =0 ,
\ee
we obtain
\be
H(x,x')={a^{n+3}\over  n^3\mu}\int
d^3\bar{\omega}\sqrt{g_\inds{\bar{\omega}}}\,{\mathbb G}_\inds{\hat{O}}(\chi,\gamma) .
\ee
The spherical symmetry of the $\bar{\omega}$ allows one to choose the coordinates on this sphere
such that $\chi$ depends only on the angle $\sigma$ on the sphere.
In these coordinates for any function  $f(\chi)$
\be
\int
d^3\bar{\omega}\sqrt{g_\inds{\bar{\omega}}}\,f(\chi)=4\pi\int_0^{\pi}d\sigma\,\sin^2\sigma\,f(\chi)
=2\pi\int_0^{2\pi}d\sigma\,\sin^2\sigma\,f(\chi) .
\ee
Thus
\be
H(x,x')=2\pi{a^{n+3}\over  n^3\mu}\int_0^{2\pi}
d\sigma\,\sin^2\sigma\,{\mathbb G}_\inds{\hat{O}}(\chi,\gamma)=
-{a^{n+1}\over  n\mu}\int_0^{2\pi}
d\sigma\,\sin^2\sigma\,{\partial\over\partial\cosh\chi}\bar{\mathbb G}(\chi,\gamma) .
\ee

Thus one can formally express the static Green function for the Maxwell field in terms of the 
scalar Green 
function ${\mathbb G}(\chi,\gamma)$ of the scalar field, which has been calculated in 
\cite{Frolov:2014kia,Frolov:2014fya}.
\be\label{G_00}
G_{00}={\mu^2(\rho^2-1)(\rho'{}^2-1)\over(M+\mu\rho)(M+\mu\rho')}{a^{n+1}\over
n\mu}\int_0^{2\pi}d\sigma\,\sin^2\sigma\,{\partial\over\partial\cosh\chi}\bar{\mathbb
G}(\chi,\gamma) .
\ee


\subsection{Even-dimensions}

In even dimensions the exact static Green function can be represented in the form 
\be\label{EvenG}
\bar{\mathbb G}(\chi,\gamma)={1\over a^{n+1}}\,{1\over 2\,(2\pi)^{n+3\over 2}}
\left({\partial\over \partial \cos\gamma}\right)^{(n+1)/2}
\,A_n  .
\ee
When $n\ge 2$, the functions $A_n(\sigma,\rho,\rho';\gamma)$ are given by the
integral
\be\begin{split}\label{A_n}
A_n&=\int_{\chi}^{\infty}dy\,{1\over\sqrt{\cosh\left({y}\right)-\cosh\left({
\chi}\right)}}\,
{\sinh\left({y\over n}\right)\over
\sqrt{\cosh\left({y\over n}\right)-
\cos\left({\gamma}\right)}}  .
\end{split}\ee
At large $y$ the integrand in \eq{A_n} behaves like $\exp[-y(n-1)/(2n)]$. Therefore,
\eq{A_n} is convergent for any $n\ge 2$. In the case of the four-dimensional
spacetime $(n=1)$ the integrand has to be modified to guarantee convergence of the
integral. For example, one can subtract the asymptotic of the integrand, which does
not depend on $\gamma$. Since \eq{EvenG} contains the derivative
of $A_n$ over $\gamma$, the resulting Green function does not depend on the
particular form of the subtracted $\gamma$-independent asymptotic. Thus,
for $n=1$ one can choose
\be\label{A_1}
A_1=\int_{\chi}^{\infty}dy\,{1\over\sqrt{\cosh\left({y}
\right)-\cosh\left({\chi}\right)}}
\,\left[{\sinh\left({y}\right)\over\sqrt{\cosh\left({y}\right)-\cos\left({\gamma}\right)}}
-{\sinh\left({y}\right)\over\sqrt{\cosh\left({y}\right)+1}}\right] .
\ee

Taking into account  \eq{G_00} we obtain
\be\label{G_00even}
G_{00}={\mu^2(\rho^2-1)(\rho'{}^2-1)\over(M+\mu\rho)(M+\mu\rho')}{1\over 
2\,(2\pi)^{n+3\over
2}}{1\over n\mu}\left({\partial\over \partial \cos\gamma}\right)^{(n+1)/2}\int_0^{2\pi}d\sigma
\,\sin^2\sigma\,\tilde{A}_n \,,
\ee
where
\be
\tilde{A}_n={\partial\over\partial\cosh\chi}A_n \,.
\ee
Thus we obtain
\be\begin{split}\label{tildeA_n}
\tilde{A}_n&=\int_{\chi}^{\infty}dy\,{1\over\sqrt{\cosh\left({y}\right)-\cosh\left({
\chi}\right)}}\,{\partial\over\partial y}
\left[{1\over\sinh(y)}{\sinh\left({y\over n}\right)\over\sqrt{\cosh\left({y\over n}\right)-
\cos\left({\gamma}\right)}}\right] .
\end{split}\ee
Note that the \eq{tildeA_n} remains to be valid for all $n$ including $n=1$ case.
In the latter case one has
\be\label{tildeA_1}
\tilde{A}_1=-{\partial\over \partial \cos\gamma}\,A_1\,.
\ee


\subsection{Odd-dimensions}

In odd-dimensional spacetimes we have
\be\label{OddG}
\bar{\mathbb G}(\chi,\gamma)={1\over a^{n+1}}\,{1\over\sqrt{2}\,(2\pi)^{{n+4\over2}}}
\left({\partial\over \partial
\cos\gamma}\right)^{n/2}\,\int_0^{2\pi} d\sigma\,B_n \,,
\ee
where
\be
B_n=\int_{\chi}^{\infty}dy\,{1\over\sqrt{\cosh
y-\cosh\chi}}\,{\sinh\left({y\over n}\right)\over
\cosh\left({y\over n}\right)-\cos\gamma} .
\ee
Taking into account  \eq{G_00} one can write
\be\label{G_00odd}
G_{00}={\mu^2(\rho^2-1)(\rho'{}^2-1)\over(M+\mu\rho)(M+\mu\rho')}{1\over\sqrt{2}\,
(2\pi)^{{
n+4\over2}}}{1\over n\mu}\left({\partial\over \partial
\cos\gamma}\right)^{n/2}\int_0^{2\pi}d\sigma\,\sin^2\sigma\,\tilde{B}_n \,.
\ee
where
\be
\tilde{B}_n={\partial\over\partial\cosh\chi}B_n \,,
\ee
\be
\tilde{B}_n=\int_{\chi}^{\infty}dy\,{1\over\sqrt{\cosh
y-\cosh\chi}}\,{\partial\over\partial y}
\left[{1\over\sinh(y)}\,{\sinh\left({y\over n}\right)\over
\cosh\left({y\over n}\right)-\cos\gamma}\right] \,.
\ee

\section{Closed form of the Green function: Examples}\label{section4}

\subsection{Four dimensions}

In four dimensions ($n=1$) the integral \eq{A_1} reads \cite{Frolov:2014kia,Frolov:2014fya}
\be
A_1=\ln\left({\cosh\left(\chi\right)+1\over
\cosh\left(\chi\right)-\cos\left(\gamma\right)} \right) .
\ee
Hence, according to \eq{tildeA_1}
\be
\tilde{A}_1=-{1\over
\cosh\chi-\cos\gamma} .
\ee

The integral over $\sigma$ in \eq{G_00even} can be taken explicitly and we obtain the closed form
for the
static Green function
\be\label{G00_4}
G_{00}(x,x')=-{1\over 4\pi 
(M+\mu\rho)(M+\mu\rho')}\left[\mu\,{\rho\rho'-\cos\gamma\over \sqrt{
\rho^2+\rho'{}^2-2\rho\rho'\cos\gamma-\sin^2\gamma}}-\mu\right] .
\ee
The last term $-\mu/[4\pi (M+\mu\rho)(M+\mu\rho')]$ in this formula describes a zero mode 
contribution, which satisfies a homogeneous
equation. One should add an extra zero mode contribution
\be
{C\over 4\pi (M+\mu\rho)(M+\mu\rho')}
\ee
with a coefficient $C$, such that the flux of the electric field across any surface surrounding the
charge and the black hole does not depend on the position of the charge.
This leads to the final result for the static Green function in four-dimensional
Reissner-Nordstr\"{o}m geometry
\be\label{G00_4}
G_{00}(x,x')=-{1\over 4\pi 
(M+\mu\rho)(M+\mu\rho')}\left[\mu\,{\rho\rho'-\cos\gamma\over \sqrt{
\rho^2+\rho'{}^2-2\rho\rho'\cos\gamma-\sin^2\gamma}}+ M\right] ,
\ee
which satisfies the correct fall-off conditions at infinity. This four-dimensional Green function 
was obtained earlier in \cite{Linet:1977vv}. The vector potential created by a point charge $e$ 
placed at the point $x'$ 
\be
J^\mu=e\,\delta(x-x')\delta_0^\mu
\ee
reads
\be
A_0(x)=4\pi e\, G_{00}(x,x') .
\ee

\subsection{Five dimensions}

In five dimensions ($n=2$) the Green function \eq{G_00odd} takes the form

\be
B_2=\int_{\chi}^{\infty}dy\,{1\over\sqrt{\cosh
y-\cosh\chi}}\,{\sinh\left({y\over 2}\right)\over
\cosh\left({y\over 2}\right)-\cos\gamma}  .
\ee
\be
B_2={\sqrt{2}\over 2}{1\over
(\cosh^2(\chi/2)-\cos^2\gamma)^{1/2}}\left[\arctan\left({
\cos\gamma\over\sqrt{\cosh^2(\chi/2)-\cos^2\gamma}}\right)+{\pi\over 2}\right] .
\ee

The Green function is
\be\label{G_00_5}
G_{00}={\mu(\rho^2-1)(\rho'{}^2-1)\over(M+\mu\rho)(M+\mu\rho')}{1\over
2\sqrt{2}\,(2\pi)^3}\left({\partial\over \partial
\cos\gamma}\right)\int_0^{2\pi}d\sigma\,\sin^2\sigma\,\tilde{B}_2 \,.
\ee
where
\be
\tilde{B}_2=\int_{\chi}^{\infty}dy\,{1\over\sqrt{\cosh
y-\cosh\chi}}\,{\partial\over\partial y}
\left[{1\over\sinh(y)}\,{\sinh\left({y\over 2}\right)\over
\cosh\left({y\over 2}\right)-\cos\gamma}\right] \,.
\ee
\be\begin{split}
\tilde{B}_2&=-{\sqrt{2}\over 4}{1\over
(\cosh^2(\chi/2)-\cos^2\gamma)^{3/2}}\left[\arctan\left({
\cos\gamma\over\sqrt{\cosh^2(\chi/2)-\cos^2\gamma}}\right)+{\pi\over 2}\right]\\
&-{\sqrt{2}\over
4}{\cos\gamma\over\cosh^2(\chi/2)(\cosh^2(\chi/2)-\cos^2\gamma)} .
\end{split}\ee

The result of integration over $\sigma$ in the \eq{G_00_5} 
\be
\int_0^{2\pi}d\sigma\,\sin^2\sigma\,\tilde{B}_2={4\sqrt{2}\,\pi\over(\rho^2-1)(\rho'{}^2-1)}\,Q
\ee
can be expressed in terms of the elliptic integrals. The result reads
\be\label{G_5}
G_{00}={\mu\over 4\,\pi^2\,(M+\mu\rho)(M+\mu\rho')}{\partial\over 
\partial\cos\gamma}\,Q ,
\ee
where
\be\begin{split}\label{G_5a}
Q&=-q\left[\mathbf{E}(\eta,\varkappa)-2\vartheta(\cos\gamma)\mathbf{E}(\varkappa)\right]\\
&+{q^2+p^2\over 
2\,q}\left[\mathbf{F}(\eta,\varkappa)-2\vartheta(\cos\gamma)\mathbf{K}(\varkappa)\right]
+\,{q_0-p_0\over q_0}\cos\gamma .
\end{split}\ee
Here $\vartheta(x)$ is the Heaviside step function and
\be\begin{split}\label{G_5b}
p&=\sqrt{\rho\rho'-\sqrt{\rho^2-1}\sqrt{\rho'{}^2-1}+1-2\cos^2\gamma}\,\big/\sqrt{2}  ,\\
q&=\sqrt{\rho\rho'+\sqrt{\rho^2-1}\sqrt{\rho'{}^2-1}+1-2\cos^2\gamma}\,\big/\sqrt{2}  ,\\
p_0&=\sqrt{\rho\rho'-\sqrt{\rho^2-1}\sqrt{\rho'{}^2-1}+1}\,\big/\sqrt{2}  ,\\
q_0&=\sqrt{\rho\rho'+\sqrt{\rho^2-1}\sqrt{\rho'{}^2-1}+1}\,\big/\sqrt{2}  ,\\
\varkappa&={\sqrt{q^2-p^2}\over q}\hh \sin\eta={q\over q_0}\mbox{sign}(\cos\gamma)  .
\end{split}\ee
Note that in spite the appearance of the Heaviside step function $\vartheta(\cos\gamma)$ in the 
expressions \eq{G_5a} the function $Q$ is continuous and smooth at 
$\gamma=Pi/2$, so that the Green function \eq{G_5} is also continuous and smooth everywhere.

On has to add to $G_{00}$ a zero mode contribution of the form
\be\label{C}
{C(\rho')\over 4\pi^2 (M+\mu\rho)} ,
\ee
in order to satisfy the boundary condition at 
infinity, meaning that the total flux of the electric field through the surface surrounding the 
electric charge should not depend on the position of the charge. 
This condition uniquely fixes the function $C(\rho')$.

When $\rho \rightarrow\infty$  we get
\be
{\partial\over 
\partial\cos\gamma}\,Q \big|_{\rho \rightarrow\infty}= 1-\rho' \,.
\ee
Therefore, one has to add the zero mode \eq{C} with
\be
C=-{M+\mu\over M+\mu\rho'}
\ee
to get a proper asymptotic of the Green function at infinity
\be
G_{00}\big|_{\rho \rightarrow\infty}=-{1\over 4\pi^2 r^2}=-{1\over 4\pi^2 
(M+\mu\rho)} .
\ee

Finally we obtain the closed for for the static Green function, satisfying the correct fall-off 
conditions at infinity
\be\label{G_5}
G_{00}=-{1\over 4\,\pi^2\,(M+\mu\rho)(M+\mu\rho')}\left[M+\mu\,{p_0\over q_0}-\mu 
{\partial\over 
\partial\cos\gamma}\,W\right] ,
\ee
where
\be\label{P}
W=-q\left[\mathbf{E}(\eta,\varkappa)-2\vartheta(\cos\gamma)\mathbf{E}(\varkappa)\right]
+{q^2+p^2\over 
2\,q}\left[\mathbf{F}(\eta,\varkappa)-2\vartheta(\cos\gamma)\mathbf{K}(\varkappa)\right] ,
\ee
and the other parameters are defined by \eq{G_5b}. This closed form for the static Green function 
of the Maxwell field is new. The vector potential The vector potential created by a point charge 
$e$ 
placed at the point $x'$ 
\be
J^\mu=e\,\delta(x-x')\delta_0^\mu
\ee
is equal to
\be
A_0(x)=4\pi e\, G_{00}(x,x') .
\ee
Here coefficient $4\pi$ comes from the normalization of the charge in the higher-dimensional 
Maxwell equations \eq{Maxwell}.


\section{Near-horizon limit}

In the vicinity of the horizon the gravitational field becomes approximately homogeneous. In the 
limit of an infinite gravitational radius the static Green function should reproduce the result 
\cite{Frolov:2014gla} for the Green function in the Rindler spacetime up to the zero mode 
contribution, because 
the topology of the Rindler horizon differs from that of the black hole horizon. One should also 
take into account that for the black hole the Killing vector is normalized to unity at infinity, 
while in the Rindler spacetime it is usually normalized to unity at the position of an 
accelerated observer, located close to the horizon.

A near-horizon limit can be derived by using the expressions
\be\begin{split}
\rho&=1+{n^2\over 2 r_\ins{g}^2}\,z^2+O(r_\ins{g}^{-4})\hhh 
t={a\over \kappa}\tilde{t}\hhh
\gamma={1\over r_\ins{g}}|\boldsymbol{x}_{\perp}-\boldsymbol{x}'_{\perp}|+O(r_\ins{g}^{-3}) \hhh
\kappa={n\mu \over r_\ins{g}^{n+1}},
\end{split}\ee
and then taking the limit $r_\ins{g}\rightarrow\infty$, while the parameter $a$ is kept finite.
The parameter $a$ has a meaning of a proper acceleration of an observer. In the limit, when the 
size of the black hole goes to infinity, the region in the vicinity of the observer is 
described by a homogeneous gravitational field, i.e., by the Rindler spacetime. The Rindler time 
coordinate $\tilde{t}$ is chosen in such a way that the timelike Killing vector 
$\xi^\alpha=\delta^\alpha_{\tilde{t}}$ has a unit norm at the position of an observer $z=a^{-1}$.

Then the metric \eq{R-N} near the horizon takes the form 
\be
ds^2=-a^2\,z^2\,d\tilde{t}^2+dz^2+d\boldsymbol{x}_{\perp}^2 + O(r_\ins{g}^{-2}).
\ee
The static Green function, corresponding to the rescailed time $\tilde{t}$ 
is $G_{\tilde{t}\tilde{t}}=\lim_{r_\ins{g}\rightarrow\infty}\,(a/\kappa)\,G_{00}$. 
Let us introduce the notations 
\be
R=\sqrt{(z-z')^2+|\boldsymbol{x}_{\perp}-\boldsymbol{x}'_{\perp}|^2} \hhh
\bar{R}=\sqrt{(z+z')^2+|\boldsymbol{x}_{\perp}-\boldsymbol{x}'_{\perp}|^2}\hhh
\zeta={\bar{R}\over 2\sqrt{zz'}}.
\ee

Then in four dimensions we get
\be\label{tildeG}
G_{\tilde{t}\tilde{t}}=-{a\over 8\pi} {R^2+\bar{R}^2\over R\,\bar{R}}-{M a\over 
4\pi\mu} .
\ee
where $G_{\tilde{t}\tilde{t}}$ is the static Green function corresponding to the rescailed time 
coordinate $\tilde{t}$. The last term in \eq{tildeG} is constant. It comes from the zero mode 
contribution. In any case this constant is a pure gauge. The boundary 
conditions at infinity of the black hole and in the Rindler spacetime are different, therefore, 
it's not surprising that zero 
mode contributions may also differ in the near-horizon limit (see discussion in 
\cite{Frolov:2014gla}). 

Similarly in five dimensions in the near-horizon asymptotic we obtain
\be\begin{split}
G_{\tilde{t}\tilde{t}}&=-{a\sqrt{zz'}\over 4\pi^2} \left[
{R^2+\bar{R}^2\over R^2\,\bar{R}^2}\mathbf{E}\left(\arcsin{\zeta},{1\over\zeta}\right)
-{1\over\bar{R}^2}\mathbf{F}\left(\arcsin{\zeta},{1\over\zeta}\right)\right]
-{a\over 4\pi^2\kappa r_\ins{g}^2}\\
&=-{3\,a\, z^2z'{}^2\over 8\pi}{1\over R^5} F\left({5\over 2}, {3\over 2}; 3; 
-{4zz'\over R^2}\right)-{a\over 4\pi^2\kappa r_\ins{g}^2}.
\end{split}\ee
When expressed in terms of the hypergeometric function, this formula exactly reproduces eq.(3.29) 
of the paper \cite{Frolov:2014gla}, where the static Green function in a homogeneous 
gravitational field was derived. The last term here is constant and can be omitted, 
because it's a pure gauge.


\section{Discussion}\label{section5}

In this paper we found the relation between static
solutions of the Maxwell field equation on the background of the $D$-dimensional 
Reissner-Nordstr\"{o}m black hole and on the background of the $(D+2)$-dimensional homogeneous 
Bertotti-Robinson 
spacetime. Using the heat kernel technique we obtained a useful integral representation for
the electric potential 
created by a 
point static charge in the Bertotti-Robinson spacetime and, hence, in the Reissner-Nordstr\"{o}m 
spacetime too. The method is very similar to the method of biconformal transformations 
\cite{Frolov:2014kia,Frolov:2014fya}, where 
the Green function and the potential created by static scalar charges near 
Reissner-Nordstr\"{o}m black holes have been calculated.

In four- and five-dimensional cases we obtained the exact static Green functions in the closed form
\eq{G00_4} and \eq{G_5}-\eq{P}. In four dimensions it correctly reproduces the well-known result 
\cite{Linet:1977vv}. To the best of our knowledge the closed form for the five-dimensional 
static Green function is new. As a test of the obtained results, we demonstrated 
that the derived static Green
functions in a generic Reissner-Nordstr\"{o}m spacetime obey a correct near-horizon limit 
\cite{Frolov:2014gla}. The obtained integral representation and analytical expressions for 
exact Green functions can be used 
to study the problem of the self-energy and self-force of point electric charges in the 
background
of higher dimensional static black holes. An interesting observation is that the self-force and 
the self-energy of charged particles qualitatively differ in odd and even spacetime dimensions 
\cite{Beach:2014aba,Frolov:2014gla,Frolov:2012xf,Frolov:2012ip,Frolov:2013qia}. In odd 
dimensions the self-force and the self-energy of point charges contain terms logarithmic in the 
distance to the horizon. These terms are related to the biconformal anomalies (see 
discussion in \cite{Frolov:2012xf,Frolov:2012ip,Frolov:2013qia}).


\acknowledgments{
This work was partly supported  by  the Natural Sciences and Engineering
Research Council of Canada. The authors are also grateful to the Killam Trust
for its financial support.}




\providecommand{\href}[2]{#2}\begingroup\raggedright\endgroup


\end{document}